\documentstyle[stwol,epsfig]{article}

\input{psfig}

\def\Journal#1#2#3#4{{#1} {\bf #2}, #3 (#4)}

\def\PLB{{\em Phys. Lett.}  B}

\def\PRD{{\em Phys. Rev.} D}

\def\ra{\rightarrow}

\def\be{\begin{equation}}
\def\ee{\end{equation}}
\def\tmt{\times 10^{-2}}
\def\tmth{\times 10^{-3}}
\def\tmf{\times 10^{-4}}
\def\tmfv{\times 10^{-5}}
\newcommand{\beq}{\begin{equation}}
\newcommand{\eeq}{\end{equation}}
\newcommand{\bea}{\begin{eqnarray}}
\newcommand{\eea}{\end{eqnarray}}
\newcommand{\barr}{\begin{array}}
\newcommand{\earr}{\end{array}}
\newcommand{\bc}{\begin{center}}
\newcommand{\ec}{\end{center}}
\newcommand{\btab}{\begin{tabular}}
\newcommand{\etab}{\end{tabular}}
\newcommand{\gv}{\mbox{GeV}}
\newcommand{\tv}{\mbox{TeV}}

\newcommand{\nn}{\nonumber}

\newcommand{\dro}{\Delta\rho}
\newcommand{\drqcd}{\delta\!\rho\,_{QCD}}
\newcommand{\roro}{\rho^{(2)}}

\newcommand{\al}{\alpha}

\newcommand{\G}{\Gamma}
\newcommand{\Gmu}{G_{\mu}}
\newcommand{\amu}{a_{\mu}}

\newcommand{\ganu}{\gamma_{\nu}}

\newcommand{\gafi}{\gamma_5}

\newcommand{\noi}{\noindent}

\newcommand{\epm}{e^+e^-}

\newcommand{\sm}{standard model }

\newcommand{\dal}{\Delta\alpha}

\newcommand{\mz}{M_Z^2}
\newcommand{\mw}{M_W^2}

\newcommand{\Dr}{\Delta r}

\newcommand{\alr}{A_{LR}}
\newcommand{\afb}{A_{FB}}

\newcommand{\ass}{asymmetries }

\newcommand{\pr}{{\it Phys.\ Rev.\ }}
 \newcommand{\prd}{{\it Phys.\ Rev.\ }{\bf D }}
\newcommand{\zp}{{\it Z.\ Phys.\ }{\bf C }}
\newcommand{\plb}{{\it Phys.\ Lett.\ }{\bf B }}
 \newcommand{\prl}{{\it Phys.\ Rev.\ Lett.\ }}
\newcommand{\np}{{\it Nucl.\ Phys.\ }{\bf B }}

\newcommand{\ms}{\overline{MS}}

\begin{document}

\title{REVIEW OF ELECTROWEAK THEORY}

\author{ W. HOLLIK }

\address{Institut f\"ur Theoretische Physik, Universit\"at Karlsruhe,
Kaiserstr.~12, D-76128 Karlsruhe, Germany}

%\author{ A.N. OTHER }

%\address{Department of Physics, Theoretical Physics, 1 Keble Road,
%Oxford OX1 3NP, England}

%%%%%%%%%%%%%%%%%%%%%%%%%%%%%%%%%%%%%%%%%%%%%%%%%%%%%%%%%%%%%%
% You may repeat \author \address as often as necessary      %
%%%%%%%%%%%%%%%%%%%%%%%%%%%%%%%%%%%%%%%%%%%%%%%%%%%%%%%%%%%%%%

\twocolumn[\maketitle\abstracts{
This talk summarizes the status of the standard model predictions for 
electroweak precision observables
including $g-2$ for muons, and discusses the status of the standard model 
and the MSSM in view of the new data.}]

\section{Introduction}
Impressive experimental results have been reported at this conference on the
$Z$ boson parameters \cite{blondel}, the $W$ mass
\cite{wmass}, and the 
top quark mass determination \cite{top,top1} with
$m_t = 175 \pm 6$ GeV. 
 
Also a sizeable amount of theoretical work has
contributed over the last few years to a steadily rising
improvement of the standard model predictions
(for a review see ref.\ \cite{yb95}). The availability of both 
highly accurate measurements and theoretical predictions, at the level
of nearly 0.1\% precision,
 provides
tests of 
the quantum structure of the standard model thereby
probing its empirically yet untested sector, and simultaneously accesses
alternative scenarios like the minimal supersymmetric extension of  
of the standard model (MSSM).

\section{Status of precision calculations}
\subsection{Radiative corrections in the standard model}

The possibility of performing precision tests is based
on the formulation of the \sm as a renormalizable quantum field
theory preserving its predictive power beyond tree level
calculations. With the experimental accuracy 
being sensitive to the loop
induced quantum effects, also the Higgs sector of the \sm
is probed. The higher order terms
induce the sensitivity of electroweak observables
to the top and Higgs mass $m_t, M_H$
and to the strong coupling constant $\al_s$.

Before one can make predictions from the theory,
a set of independent parameters has to be taken from experiment.
For practical calculations the physical input quantities
$ \al, \; \Gmu,\; M_Z,\; m_f,\; M_H; \; \al_s $
are commonly used    
for fixing the free parameters of the standard model.
 Differences between various schemes are formally
of higher order than the one under consideration.
 The study of the
scheme dependence of the perturbative results, after improvement by
resumming the leading terms, allows us to estimate the missing
higher order contributions.
 
\smallskip
Two sizeable effects in the electroweak loops deserve a special
discussion:
\begin{itemize}
\item
The light fermionic content of the subtracted photon vacuum polarization
corresponds to a QED induced shift
in the electromagnetic fine structure constant. The evaluation of the
light quark content
 \cite{eidelman}
 yield the result
\beq  (\dal)_{had} = 0.0280 \pm 0.0007\, . \eeq
Other determinations \cite{swartz}
agree within one standard deviation. Together with the leptonic
content, $\dal$ can
be resummed resulting in an effective fine structure
constant at the $Z$ mass scale:
\beq
   \al(\mz) \, =\, \frac{\al}{1-\dal}\,=\,
   \frac{1}{128.89\pm 0.09} \, .
\eeq
 \item
The electroweak mixing angle is related to the vector boson
masses  by
\bea
  \sin^2\theta  = 
   1-\frac{\mw}{\mz} + \frac{\mw}{\mz} \dro\, +  \cdots
\eea
where
the main contribution to the $\rho$-parameter 
 is from the  $(t,b)$ doublet \cite{rho},
at the present level calculated to
 \beq
 \dro= 3 x_t \cdot [ 1+ x_t \,  \roro+ \drqcd ]
\eeq
with
\beq
 x_t =
 \frac{\Gmu m_t^2}{8\pi^2\sqrt{2}} \, .
\eeq
 The electroweak 2-loop
 part \cite{bij,barbieri} is described by the
function $\roro(M_H/m_t)$.
$\drqcd$ is the QCD correction
to the leading $\Gmu m_t^2$ term
 \cite{djouadi,tarasov}
$$
    \drqcd = - 2.86 a_s  - 14.6 a_s^2, \;\;\;\;
 a_s = \frac{\al_s(m_t)}{\pi} \, .
$$

\end{itemize}
\subsection{The vector boson masses}
The correlation between
the masses $M_W,M_Z$ of the vector bosons,          in terms
of the Fermi constant $\Gmu$, is in 1-loop order given by
 \cite{sirmar}:
$$
\frac{\Gmu}{\sqrt{2}}   =
            \frac{\pi\al}{2s_W^2 M_W^2} [
        1+ \Dr(\al,M_W,M_Z,M_H,m_t) ] \, .
$$
 
\medskip \noi
The appearance of large terms in $\Dr$ requires the consideration
of higher than 1-loop effects.
At present, the following  
higher order contributions are available:
\begin{itemize}
\item
The leading log resummation \cite{marciano} of $\dal$: \\
$  1+\dal\, \ra \, (1-\dal)^{-1}$
\item
The incorporation of
non-leading higher order terms
containing mass singularities of the type $\al^2\log(M_Z/m_f)$
from the light fermions \cite{nonleading}.
\item
The resummation of the leading $m_t^2$ contribution \cite{chj}
in terms of $\dro$ in Eq.\ (4).
 Moreover, the complete
 $O(\al\al_s)$ corrections to the self energies
 are available \cite{qcd,dispersion1},
 and part of the $O(\al\al_s^2)$ terms \cite{steinhauser}.

\item
The $\Gmu^2m_t^2 M_Z^2$ contribution 
of the electroweak 2-loop order \cite{padova}.
\end{itemize}

\subsection{$Z$ boson observables}
With $M_Z$ as a precise input parameter, 
the predictions for the partial widths
as well as for the asymmetries
can conveniently be calculated in terms of effective neutral
current coupling constants for the various fermions:
\bea
 & &
 J_{\nu}^{NC}  ~   \,
  g_V^f \,\ganu -  g_A^f \,\ganu\gafi  \\
 & &
   =  \left( \rho_f \right)^{1/2}
\left( (I_3^f-2Q_fs_f^2)\ganu-I_3^f\ganu\gafi \right)  . \nn
\eea
with form factors 
$\rho_f$ and $s_f^2$ for the overall normalization and the
effective mixing angle.

\smallskip
The effective mixing angles are of particular interest since
they determine the on-resonance asymmetries via the combinations
   \beq
    A_f = \frac{2g_V^f g_A^f}{(g_V^f)^2+(g_A^f)^2}  \, .
\eeq
Measurements of the \ass hence are measurements of
the ratios
\beq
  g_V^f/g_A^f = 1 - 2 Q_f s_f^2
\eeq
or the effective mixing angles, respectively.

\smallskip
The total
$Z$ width $\Gamma_Z$ can be calculated
essentially as the sum over the fermionic partial decay widths.
Expressed in terms of the effective coupling constants they
read up to 2nd order in the fermion masses:
\bea
\Gamma_f
  & = & \G_0
 \, \left(
     (g_V^f)^2  +
     (g_A^f)^2 (1-\frac{6m_f^2}{\mz} )
                           \right)        \nn \\
 &  & \cdot   (1+ Q_f^2\, \frac{3\al}{4\pi} ) 
          + \Delta\G^f_{QCD} \nn
\eea
with
$ \left[ N_C^f = 1
 \mbox{ (leptons)}, \;\; = 3 \mbox{ (quarks)} \right] $ 
$$
\G_0 \, =\,
  N_C^f\,\frac{\sqrt{2}\Gmu M_Z^3}{12\pi},
$$
and the QCD corrections  $ \Delta\G^f_{QCD} $
 for quark final states
 \cite{qcdq}.

\subsection{Accuracy of the standard model predictions}
 For a discussion of the theoretical reliability
of the \sm predictions one has to consider the various sources
contributing to their
uncertainties:

The experimental error of the hadronic contribution
to $\al(\mz)$, Eq.\ (2), leads to
$\delta M_W = 13$ MeV in the $W$ mass prediction, and
$\delta\sin^2\theta = 0.00023$ common to all of the mixing
angles, which matches with the experimental precision.

The uncertainties from the QCD contributions
can essentially be traced back to
those in the top quark loops for the $\rho$-parameter.
They  can be combined into the following errors
\cite{kniehl95}:
$$
 \delta(\dro) \simeq 1.5\cdot 10^{-4},   \;
 \delta s^2_{\ell} \simeq 0.0001 \, .
$$

The size of unknown higher order contributions can be estimated
by different treatments of non-leading terms
of higher order in the implementation of radiative corrections in
electroweak observables (`options')
and by investigations of the scheme dependence.
Explicit comparisons between the results of 5 different computer codes  
based on  on-shell and $\ms$ calculations
for the $Z$ resonance observables are documented in the ``Electroweak
Working Group Report'' \cite{ewgr} in ref.\ \cite{yb95}.
Table 1  shows the uncertainty in a selected set of
precision observables.
Quite recently (not included in table 1)
 the non-leading 2-loop corrections
$\sim \Gmu^2m_t^2 M_Z^2$ have been calculated \cite{padova}
for $\Delta r$ and $s_{\ell}^2$.
They reduce the uncertainty in $M_W$ and $s^2_{\ell}$ considerably,
by about a factor 0.2.

\begin{table}[htbp]\centering
\caption[]
{Largest half-differences among central values $(\Delta_c)$ and among
maximal and minimal predictions $(\Delta_g)$ for $m_t = 175\,\gv$,
$60\,\gv < M_H < 1\,\tv$, $\al_s(\mz) = 0.125$
(from ref.\ \cite{ewgr}) }
\vspace{0.5cm}
\begin{tabular}{c|c|c}
\hline 
Observable $O$ & $\Delta_c O$  & $\Delta_g O$ \\
\hline
            & & \\
$M_W\,$(GeV)          & $4.5\tmth$ & $1.6\tmt$\\
$\G_e\,$(MeV)          & $1.3\tmt$ & $3.1\tmt$\\
$\G_Z\,$(MeV)          & $0.2$     & $1.4$\\
$ s^2_e$             & $5.5\tmfv$ & $1.4\tmf$\\
$ s^2_b$             & $5.0\tmfv$ & $1.5\tmf$\\
$R_{had}$                 & $4.0\tmth$& $9.0\tmth$\\
$R_b$                 & $6.5\tmfv$ & $1.7\tmf$ \\
$R_c$                 & $2.0\tmfv$& $4.5\tmfv$ \\
$\sigma^{had}_0\,$(nb)    & $7.0\tmth$ & $8.5\tmth$\\
$\afb^l$             & $9.3\tmfv$ & $2.2\tmf$\\
$\afb^b$             & $3.0\tmf$ & $7.4\tmf$ \\
$\afb^c$             & $2.3\tmf$ & $5.7\tmf$ \\
$\alr$                & $4.2\tmf$ & $8.7\tmf$\\
\hline 
\end{tabular}
 
%\label{ta9}
\end{table}
%\normalsize

\section{Standard model and precision data}
In table 2
the \sm predictions for $Z$ pole observables and the $W$ mass  are
put together for a light and a heavy Higgs particle with $m_t=175$ GeV.
 The last column is the variation of the prediction according to
$\Delta m_t = \pm 6$ GeV. The input value $\al_s = 0.123$
is the one from QCD observables at the $Z$ peak \cite{alfas}.
Not included are the uncertainties from
$\delta\al_s=0.006$, which amount to 3 MeV for the hadronic $Z$ width.
The experimental results on the $Z$ observables are from combined 
LEP and SLD data. $\rho_{\ell}$ and $s^2_{\ell}$ are the leptonic
neutral current couplings in eq.\ (6), obtained from partial widths and
asymmetries  under the assumption of lepton universality.
Compared to the previous year, the deviation in $R_c$ has disappeared
and $R_b$ is on its way towards the standard model value.
On the other hand, the deviation in $A_b$ has become somewhat stronger.

Table 2 also illustrates the sensitivity of the various quantities 
to the Higgs mass.
The effective mixing angle turns out to be
the most sensitive  observable, where both the experimental error and the
uncertainty from $m_t$ are small compared to the variation with $M_H$.
Since a light Higgs boson corresponds to a low value of $s^2_{\ell}$,
the strongest upper bound on $M_H$ is from $A_{LR}$ at the SLC \cite{sld},
 whereas 
LEP data alone allow to accommodate also a relatively heavy Higgs.

%\newpage
\begin{table*}[t]
            \caption{Precision observables: experimental results 
             {\protect\cite{blondel}}
             and standard model         
             predictions. } \vspace{0.5cm}
            \bc
 \btab{| l | l | r | r | r | }
\hline
 observable & exp. (1996) &  $M_H=65$ GeV & $M_H=1$ TeV &
 $ \Delta m_t $ \\
\hline
$M_Z$ (GeV) & $91.1863\pm0.0020$ &  input & input &    \\
\hline
$\Gamma_Z$ (GeV) & $2.4946\pm 0.0027$ & 2.5015 & 2.4923 & $\pm 0.0015$ \\
%\hline
%$\Gamma_{had}$ (GeV) & $1.740\pm 0.008$ &
% $1.736\pm 0.008 \pm 0.007$ \\
%\hline
%$\Gamma_e$ (MeV) & $83.2\pm 0.4$ & $83.7\pm 0.4 $ \\
\hline
$\sigma_0^{had}$ (nb) & $41.508\pm 0.056$ & 41.441 & 41.448 & $\pm 0.003$  \\
\hline
 $\G_{had}/\G_e$ & $20.778\pm 0.029 $ & 20.798 & 20.770 & $\pm 0.002$ \\
%\hline
%$\Gamma_e$ (MeV) & $83.82\pm 0.27$ & $83.7\pm 0.4 $ \\
%\hline
%$\Gamma_{inv}$ (MeV) & $499.9\pm 2.5$ & $501.6\pm 1.1$ \\
\hline
$\G_b/\G_{had}=R_b$  & $0.2178\pm 0.0011$ & 0.2156 & 0.2157 & $\pm 0.0002$ \\
\hline
$\G_c/\G_{had}=R_c$  & $0.1715\pm0.0056$ & 0.1724 & 0.1723 & $\pm 0.0001$ \\
\hline
$A_b$            & $0.867\pm 0.022$  & 0.9350 & 0.9340 & $\pm 0.0001$  \\
\hline
$\rho_{\ell}$ & $1.0043\pm 0.0014$ & 1.0056 & 1.0036 & $\pm 0.0006$ \\
\hline
$s^2_{\ell}$  & $0.23165\pm 0.00024$ & 0.23115 & 0.23265 & $\pm 0.0002$ \\
%\hline
%$s^2_e (A_{LR})$ & $0.23049\pm 0.00050$ & $0.2317\pm 0.0012$   \\
% LEP$+$SLC   &  $0.23143\pm 0.00028$    &                    \\
\hline
$M_W$ (GeV) & $80.356 \pm 0.125$ & 80.414 & 80.216 & $\pm 0.038$  \\
\hline
\etab
\ec 
%  \vspace{-1.5cm}
\clearpage
\end{table*}

\paragraph{\it Standard model fits and Higgs mass range:}
Assuming the validity of the \sm a global fit to all electroweak results from
LEP, SLD, $M_W$,  $\nu N$ and $m_t$, 
allows to derive information on the allowed range for
the Higgs mass. 
The impact of $R_b$
 is only marginal,  whereas $A_{LR}$
is decisive for a restrictive upper bound for $M_H$.
From the fit one obtains \cite{blondel,deboer}:
% \cite{higgs95}):   
$$
  M_H = 149 ^{+148}_{-82} \gv, \;\;\;
  M_H < 450 \gv (95\% C.L.)  \, .
$$
Similar results have been obtained in \cite{jellis} 
(updated from \cite{higgs95}).
Without $\alr$,  the 95\% C.L upper bound is shifted upwards by about
260 GeV.

\smallskip \noi
These numbers do not yet include the theoretical uncertainties of the
standard model predictions. The LEP-EWWG \cite{blondel,gruenewald}
has performed a study of the influence of the various `options'
(section 2.4) on the Higgs mass with the result
that the 95\% C.L. upper bound is shifted by +100 GeV to higher values.
It has to be kept in mind, however, that this error estimate is based on the
uncertainties as given in table 1. Since the recent improvement in the 
theoretical prediction \cite{padova}
is going to reduce the theoretical uncertainty
especially in the effective mixing angle one may expect also a significantly
smaller theoretical error on the Higgs mass bounds once the 2-loop terms
$\sim \Gmu^2 m_t^2 \mz$ are implemented in the codes used for the fits.
At the present stage the codes are without the new terms.

\section{Fermion pair production above the Z resonance}
Also above the $Z$ peak, the production of fermion pairs is an
important class of processes since they are the dominant ones at LEP 2.
The cross sections  are measurable with high accuracy: 1.2\% for
$\epm \ra \mu^+\mu^-$, and 0.7\% for $\epm \ra hadrons$ \cite{boudjema}.
From the theoretical side they are of special interest because box diagrams
with two heavy boson exchange are no longer negligible, contributing
several percent to the integrated cross section. 
Large QED corrections from the radiative tail of the $Z$ resonance
to both integrated cross sections and forward-backward asymmetries 
occur and require a careful theoretical treatment to obtain a 
theoretical precision of about 0.5\%
(see \cite{boudjema} for more information).

\section{Muon anomalous magnetic moment}
The anomalous magnetic moment of the muon,
\beq
   a_{\mu} = \frac{g_{\mu}-2}{2}
\eeq
provides a precision test of the standard model at low energies.
Within the present experimental accuracy of
$\Delta\amu = 840\cdot 10^{-11}$, theory and experiment are in best
agreement, but the electroweak loop corrections are still hidden
in the noise. A new experiment, E 821 at Brookhaven National
Laboratory \cite{brookhaven}, is being prepared for 1997 to reduce
the experimental error down to $40\pm 10^{-11}$ and hence will
become sensitive to the electroweak loop contribution.

For this reason the standard model prediction has to be known with
comparable precision. Recent theoretical work has contributed to
reduce the theoretical uncertainty by calculating the electroweak
2-loop terms \cite{ew22,ew23}
and updating the contribution from the hadronic
photonic vacuum polarization
(first reference of \cite{eidelman})
$$
 \amu^{had}(\mbox{vacuum pol.}) = (7024\pm 153)\cdot 10^{-11}
$$
which agrees within the error with the result of \cite{dub}.
The
main sources for the theoretical error at present are the hadronic
vacuum polarization and the light-by-light scattering mediated by
quarks, as part of the 3-loop hadronic contribution
\cite{sanda,bijnens}.
Table 3 shows  the breakdown of $\amu$. The hadronic part
is supplemented by the higher order $\al^3$
vacuum polarization effects \cite{had3} but is without the
light-by-light contribution.

\begin{table}[htbp]\centering
\caption[]
{Contributions $\Delta\amu$ to the muonic anomalous magnetic moment
and their theoretical uncertainties, in units of $10^{-11}$.  }
\vspace{0.5cm}
\begin{tabular}{l|r|r}
\hline 
source & $\Delta\amu$  & error \\
\hline
            & & \\
QED \cite{qed} & 116584706 & 2  \\
hadronic \cite{eidelman,had3}  & 6916  & 153 \\
EW, 1-loop \cite{ew1} & 195 &     \\
EW, 2-loop \cite{ew23}  & -44   &  4  \\
light-by-light \cite{sanda} & -52 &  18 \\
light-by-light \cite{bijnens} & -92 &  32 \\
\hline
future experiment &   & 40  \\
\hline 
\end{tabular}
 
%\label{ta9}
\end{table}
%\normalsize
 
The 2-loop electroweak contribution is as big in
size as the expected experimental error.
The dominating theoretical uncertainty at present is the error in
the hadronic vacuum polarization. But also the contribution involving
light-by-light scattering needs improvement in order to reduce the
theoretical error.

\section{The MSSM and precision data}
The MSSM deserves a special discussion
as the most predictive framework beyond the minimal model.
Its structure allows a similarly complete calculation of
the electroweak precision observables
as in the standard model in terms of one Higgs mass
(usually taken as $M_A$) and $\tan\beta= v_2/v_1$,
together with the set of
SUSY soft breaking parameters fixing the chargino/neutralino and
scalar fermion sectors.
It has been known since quite some time
\cite{higgs}
that light non-standard
Higgs bosons as well as light stop and charginos
% all around 50 GeV or little higher,
predict larger values for the ratio $R_b$ \cite{susy1,susy3}.
Complete 1-loop calculations are available for
$\Delta r$ \cite{susydelr} and for the $Z$ boson observables
\cite{susy3}.

For obtaining the optimized SUSY parameter set a global
fit to all the electroweak precision data, including the top
mass measurement,
has been performed with the new data \cite{deboer}.
Figure 1 displays the
experimental data normalized to the best fit results  
in the SM and MSSM, with 
the data from this conference.
The difference between the experimental and the SM value
of $R_b$ can now be fully explained by the MSSM. Other quantities are
practically unchanged.
In total, the $\chi^2$ of the fit is slightly better than in the
standard model, but due to the larger numbers of parameters, the
probability for the standard model is higher.
A similar situations occurs for large $\tan\beta$ with light
Higgs bosons $h^0,A^0$ around 50 GeV.

\setlength{\unitlength}{0.7mm}
\begin{figure}[hbt]
\mbox{\epsfxsize8.0cm\epsffile{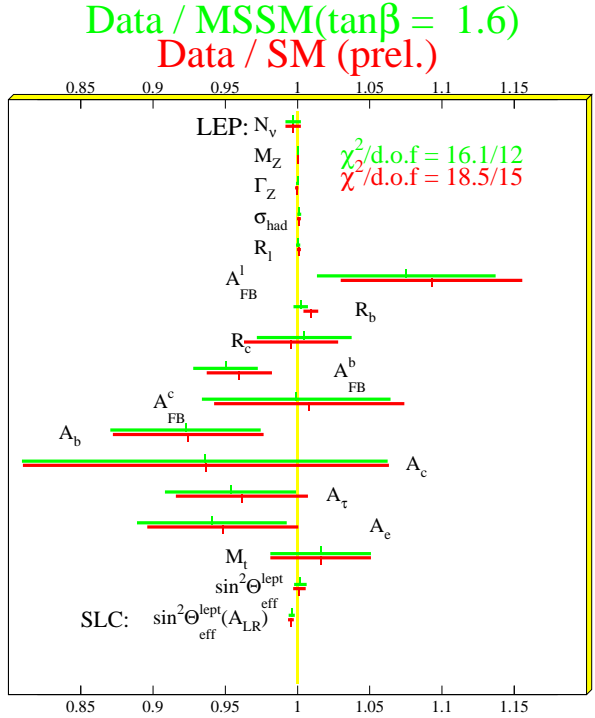}}
\vspace{-1.5cm}
\caption{Experimental data normalized to the best fit results in
         the SM and MSSM.}
\end{figure}

\bigskip
In conclusion, the theoretical predictions for electroweak precision
observables are reliably calculated, with the main uncertainty from $\dal$
in (2).
In view of the new data,
the standard model is in a very good shape.
The MSSM is competitive to the standard model, but it is no longer prefered
by the data.

\bigskip \noi
{\bf Acknowledgements:}
I want to thank W. de Boer, P. Gambino, M. Gr\"unewald, G. Passarino,
U. Schwickerath and G. Weiglein for helpful discussions and valuable
informations.

\end{document}